# Wideband impedance measurement techniques in small complex cavities such as ear simulators and the human ear canal

Søren Jønsson, Andreas Schuhmacher, Henrik Ingerslev

Brüel & Kjær Sound & Vibration A/S, Skodsborgvej 307, DK-2850, Denmark

ABSTRACT

The multimedia evolution has led to increased audio signal bandwidth in new generations of smartphones, headsets, headphones, as well as hearing aids. Full audio bandwidth performance, up to 20 kHz, is required. This calls for wider band performance of the ear simulators, and head and torso simulators used by the industry to evaluate these multimedia devices, as well as a better understanding of the high frequency behaviour of the human ear that these simulators are supposed to replicate. Acoustic input impedance measurements are an important parameter when characterizing small complex cavities such as ear simulators and the human ear. Today, this is well covered up to around 8-10 kHz. In this study a calibration procedure for measuring wideband impedance is developed, that uses multiple reference loads each defined by a detailed simulated impedance, in order to increase the applicable frequency bandwidth of the measurements. Comparing measured impedance magnitude, calibrated with this new procedure, to detailed simulations on a widely used ear simulator (IEC711 coupler) shows good agreement in the full audio bandwidth. Furthermore, the calibration procedure is streamlined to make it suited for a larger scale impedance study of the human ear, and measurements in one human ear is presented.

# I. INTRODUCTION

The acoustic impedance in the human ear canal is closely related to transmission of sound to the middle ear and impedance measurements have been an important aid in examining the middle ear function for many years. Earlier studies[1–5] have led to the specification and requirements of ear simulators used even up until today and embodied in international standards[6–10] to test the various types of earphones in smartphones, headsets, headphones and hearing aids.

The sound pressure levels generated by typical earphones are largely affected by the ear canal impedance, since they do not approximate ideal sound pressure generators when coupled to the human ear. The sound pressure developed especially for supra-aural earphones is also affected by any leak introduced between the earphone and the ear. This can cause losses at lower frequencies, typically below 500 Hz, and some gain or variations in the mid frequency range between 600 Hz and 1.5 kHz. [11,12]

Impedance measurements of the human ear are, however, only well covered in a limited bandwidth. In the earlier studies[1–5] data are reported from about 200 Hz to 6.5 kHz at most. Even in later studies[12–19], no consistent data seems to be available at frequencies above 8-10 kHz.

The purpose of this study is to develop a reliable method to measure wideband impedance in the human ear canal. Measurements in the human ear canal are known to be quite difficult to perform, and a larger study often takes place over several days, or weeks. Several factors contribute to these difficulties. Two important ones are the condition of the subject on the day the measurement is scheduled to take place, and that measurements are usually obtained in the ear canal at some unknown distance from the ear drum. This will be addressed in a larger in-vivo impedance study on human ears[20], that uses the measurement techniques developed here.

Another important factor that will be addressed here is that at higher frequencies the wavelength of sound becomes close to the dimensions within the ear canal and concha, where the possible model assumptions made for calculating the impedance may cease to be valid. For a typical supra-aural earphone applied to a normal ear, the largest dimension behind the enclosed ear is about 40 mm from the earphone to the eardrum. At frequencies around 4.3 kHz, half a wavelength resonance will therefore occur and may affect the accuracy unless the measurement methods used for calculating the impedance can take that into account. Models based on lumped-elements analogies will only be accurate when the largest dimensions are small compared to the wavelength. For a typical insert earphone, the largest dimension from the tip of the earphone to the eardrum is only about 13 mm. Resonances therefore move up in frequency, to about 14 kHz for the half wavelength resonance. The same measurement method when applied to insert earphones will therefore produce more accurate results at higher frequencies than when applied for supra-aural earphones.

The work presented here will investigate and refine the "two-load" method [13,21–23] which is based on measurements in two known reference load cavities for estimating the Thévenin equivalent of a transducer source, or the input impedance of an unknown load connected to the source. The limitations associated with the two-load method, is that is starts to break up when the frequency approaches the first antiresonance around 5-6 kHz in a typical human ear canal, and that it tends to generate a non-physical negative resistance. To overcome the shortcomings of the two-load method impedance estimates from numerical simulations are introduced and additionally using more than two reference loads may eliminate the problem associated with estimating negative resistance for the unknown cavity. All this leads to a measurement technique developed to extend the applicable frequency range to the full audio bandwidth up to 20 kHz when analysing the input impedances of small complex cavities such as ear simulators, and the human ear canal. The calibration procedure involving more than two reference loads is further improved to reduce the calibration time, to make it more efficient to use in a large scale in-vivo study.

## II. MEASUREMENT SET-UP

Measurements for the calibration methods discussed and evaluated in this paper were done using a dedicated impedance probe and additionally a set of reference load cavities were manufactured.

A photo and a schematic of the measurement set-up is shown in Figure 1. In the upper part of Figure 1(a) the reference load cavities used in the calibration procedures are shown. The reference loads are simple cylindric cavities made in brass with volumes ranging from 40 mm$^3$ to 1285 mm$^3$. Each reference load is named according to its volume, e.g., the reference load named V759 has a volume of 759 mm$^3$, and in Table I the names and geometries of all seven reference loads are listed. In the top right of Figure 1(a) there is an ear simulator (the IEC711 coupler) representing a complex unknown load as described in Section A below.

In the lower part of Figure 1(a) the impedance probe assembly is shown. It contains two impedance matched condenser probe microphones, Brüel & Kjær Type 4182. One probe microphone is used as a high impedance receiver, and the other as a high output impedance transmitter. In this symmetrical configuration the volume displacement generated by the transmitter is nearly constant with frequency, and the sound pressure picked up by the receiver is therefore nearly proportional to the input impedance of the load multiplied by the frequency [24]. The two probe microphones are connected through two identical thin acoustic waveguide tubes to the loads. The two acoustic waveguide tubes have an inner diameter of 0.9 mm, a length of 48 mm, and a spacing of 3 mm at the measurement plane.

As illustrated in Figure 1(b) the impedance probe assembly on the left side is in turn connected to reference load cavity 1, reference load cavity 2, and an unknown load on the right side, shown here as a simple cavity, or an ear canal, or an ear simulator.

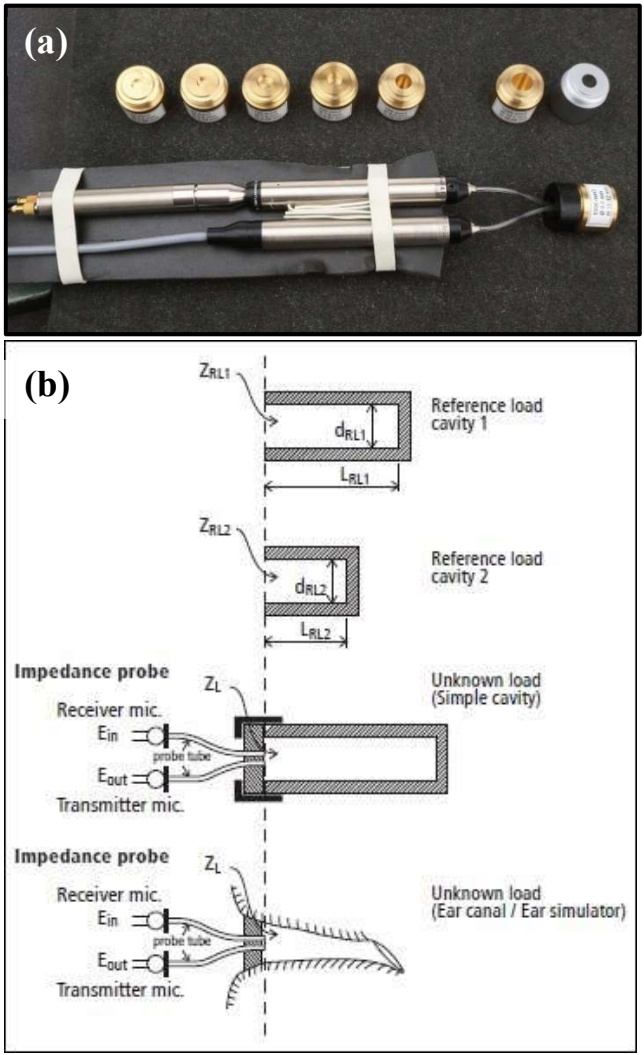

Figure 1 (a): A photo of the measurement set-up with the cylindrical reference load cavities made in brass and an ear simulator (the IEC711 coupler) in the upper part. In the lower part the impedance probe assembly with two high impedance receiver and transmitter probe microphones connected to one reference load cavity through thin acoustic waveguide tubes.

Figure 1 (b): Schematic of the measurement set-up. The impedance probe assembly on the left side is in turn coupled to reference load cavity 1, reference load cavity 2, and the unknown load on the right side, shown here as a simple cavity, or an ear canal, or an ear simulator.

Table I: Details of the seven cylindrical reference load cavities used in the calibration procedures. The reference loads are named according to their volume, e.g., V759 has a volume of 759 mm$^3$. The first mode and the impedance at 100 Hz in the last two columns are found from simulations.

| Reference load name | Volume (mm$^3$) | Diameter (mm) | Length (mm) | First mode (kHz) | Impedance at 100 Hz (GPa s/m$^3$) |
|---|---|---|---|---|---|
| **V40**   | 40   | 5.5  | 1.68  | > 25 | 5.6  |
| **V65**   | 65   | 5.5  | 2.74  | > 25 | 3.5  |
| **V100**  | 100  | 7.6  | 2.20  | > 25 | 2.3  |
| **V200**  | 200  | 7.6  | 4.41  | > 25 | 1.1  |
| **V759**  | 759  | 8.0  | 15.10 | 6.5  | 0.3  |
| **V1006** | 1006 | 9.0  | 15.81 | 6.0  | 0.22 |
| **V1285** | 1285 | 10.1 | 16.04 | 5.8  | 0.18 |

### A. The ear simulator (IEC711 coupler)

The occluded ear simulator is introduced here as an example of a rather complex load impedance (comparable to a human ear) and is described in the international standard IEC60318-4.[7] Formerly it was named the IEC711 standard and the ear simulator is often referred to as the IEC711 coupler. It is widely used for hearing aid testing of earphones coupled to the ear by ear inserts, such as tubes or ear moulds. It is also widely used as an integral part of all Type 3 ear simulators for measurements on telephones and headsets as described in ITU-T recommendation P.57 and P.58. [8,9]

A sectional drawing of the IEC711 coupler and its coupling interface is shown in Figure 2. The IEC711 coupler and the reference load cavities all have the same coupling interface (see the upper part of Figure 1 (a)). This allows for a uniform and accurate mounting of the impedance probe assembly without leak or added volume. The IEC711 coupler consists of a centre main cavity with a diameter of 7.5 mm and a length of 12.3 mm with two annular cavities connected to it through two narrow slits. The main cavity is terminated with a half-inch measuring microphone. The IEC711 coupler approximates the impedance of the human ear canal from approximately 10 mm behind its opening.

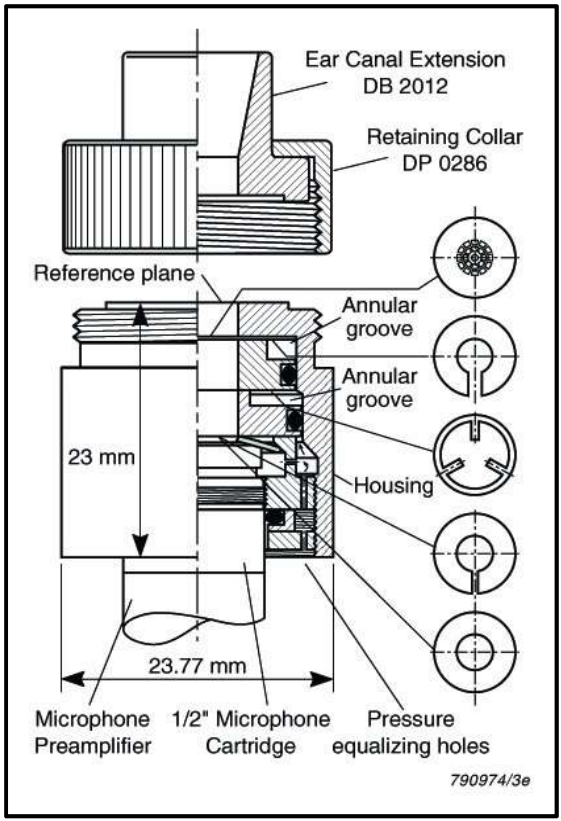

Figure 2: Sectional drawing of the ear simulator (IEC711 coupler) and its coupling interface, used here as an example of a complex load impedance comparable to a human ear.

## III.  SIMULATION

One of the improvements of the traditional two-load method introduced in this work is related to numerical simulation involving the reference load cavities.

3D models of all reference loads (see Table I) and the IEC711 coupler (see Figure 2) are generated using CAD software. The commercial software package COMSOL Multiphysics Acoustics Module is used for creating meshes and finite element modelling of all sound fields and impedances. The mesh resolution is chosen relative to the complexity of the structure, with up to 5 elements pr. mm for the IEC711 coupler, and it is continuously checked. If necessary, the mesh resolution is increased until no significant differences in the calculations are observed. Viscothermal loss effects are also included when calculating the impedance of the IEC711 coupler in order to include losses from the narrow slits between the main cavity and the two annular cavities [25].

Figure 3 shows the magnitude of the sound pressure inside the large reference load V1285, the small reference load V200, and the IEC711 coupler. The two acoustic waveguide tubes are located on top of the cylinder, the transmitter to the left and the receiver to the right. The magnitude of the sound pressure inside V1285 is shown at the quarter wavelength resonance 5.95 kHz (a) and at the half wavelength resonance 10.7 kHz (b). In reference load V200 the sound field is uniform in most parts of the cavity up to around 20 kHz. At 25 kHz the sound field starts to break up, and at 26.5 kHz the first cross resonance appears as shown in (c), which then changes to a length resonance at 38.9 kHz as shown in (d).

At frequencies below the resonances the sound field is relative uniform within the cavities, except close to the waveguide tubes. In Figure 3(a) this is seen as an increase in sound pressure around the transmitter tube. This is clearly different from a simple analytic calculation on a cavity without the waveguide tubes.

The advantage of using smaller reference loads is that undesirable effects of the resonances is moved to higher frequencies and out of the frequency range of interest, whereas the disadvantage is that they have a high impedance especially at low frequencies, often much higher than the unknown load. Hence, the larger reference loads (e.g., V759, V1006, and V1285) are best at low frequencies, and the small loads (e.g., V40, V65, V100, and V200) are best at high frequencies.

The simulated magnitude of the sound pressure inside the IEC711 coupler at the half wavelength resonance at 12.8 kHz, is shown Figure 3(e). The two acoustic waveguide tubes of the impedance probes are shown on the top of the main cavity, the transmitter to the left and the receiver to the right.

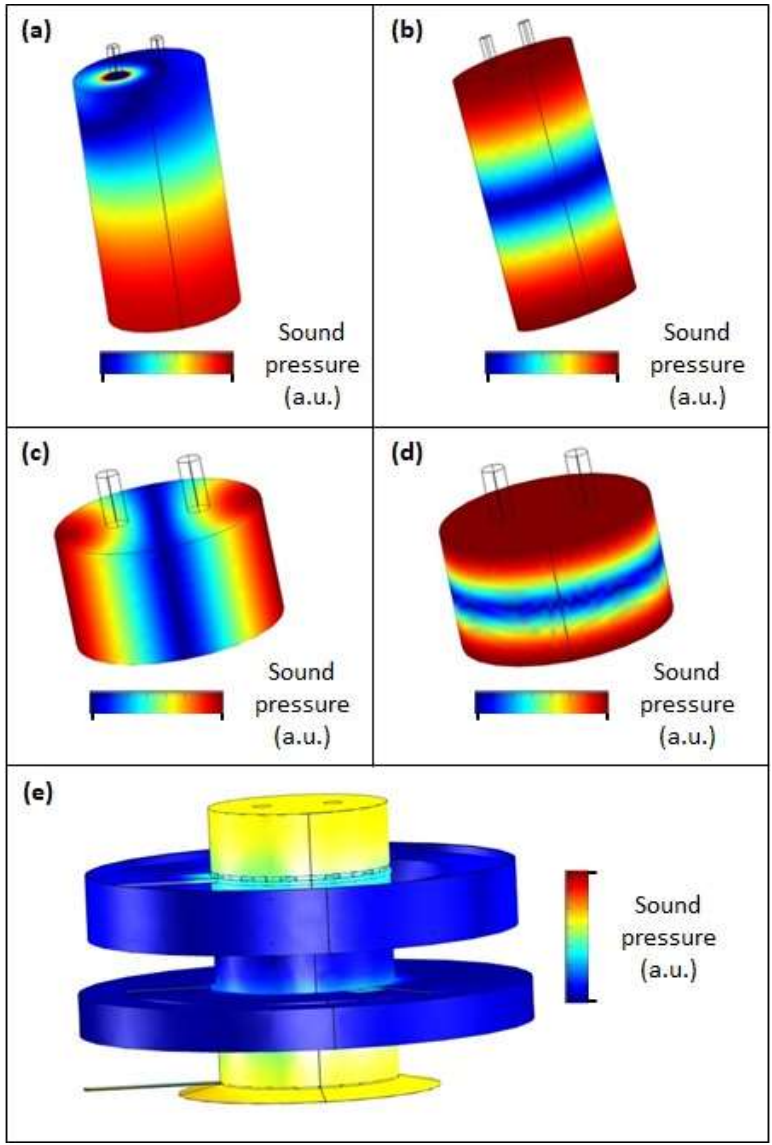

Figure 3: Simulated magnitude of the sound pressure inside the large reference load cavity of 1285 mm$^3$ (V1285), the smaller cavity of 200 mm$^3$ (V200), and of the IEC711 coupler at selected frequencies. The two acoustic waveguide tubes are located on the top of the cylinder, the transmitter to the left and the receiver to the right. (a) and (b) show the quarter and half wavelength resonances at 5.95 kHz and 10.7 kHz, respectively, for V1285. (c) and (d) show the first transverse modes at 26.5 kHz, and the first length mode at 38.9 kHz for V200. (e) shows the half wavelength resonance at 12.8 kHz for the IEC711 coupler.

For simple cylindrical cavities as the reference loads, the impedance of the fundamental mode in the radial direction can be estimated analytically by [26]:

$$Z_{RL} = \frac{-i\rho c}{A \tan\left(\frac{\omega}{c}L\right)} \quad (1)$$

where $\rho$ is the density of air, $c$ is the speed of sound in air, $\omega$ is the angular frequency, $A$ is the cross-sectional area and $L$ is the length of the cylinder. Here it is assumed that the sound pressure may be regarded as constant across the diameter of the cylinder and that sound pressure waves propagate along the longitudinal direction as plane waves. For this to be true the wavelength must be much longer than the radius of the cylinder [13].

Figure 4 shows a comparison of the impedance magnitude of reference load V1285 calculated by the analytic expression in equation ( 1 ) and using finite element simulation. At frequencies below 2-3 kHz approximately the same impedance is obtained, whereas beyond 2-3 kHz large differences are seen around the antiresonances.

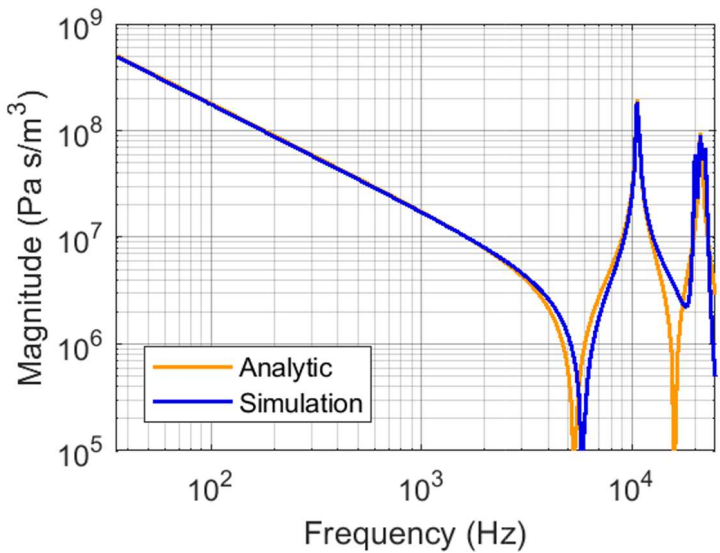

Figure 4: Comparison of the impedance magnitude of reference load cavity V1285 calculated using the analytic expression in equation ( 1 ) and by finite element simulation.

## IV. CALIBRATION PROCEDURES

The different calibration procedures are reviewed in this section, ranging from the traditional two-load method to a multi-load procedure which finally is turned into a practical calibration procedure suited for large scale studies involving many humans.

Figure 5 shows the full acoustic circuit of the measurement set-up in (a), and the Thévenin equivalent acoustic circuit in (b) [16,27]. $Z_{T1}$, and $Z_{T2}$ are the impedance of the two waveguide tubes connecting the probe microphones and the load. $Z_{M1}$, and $Z_{M2}$ are the impedance of the impedance matching tubes inside the two probe microphones. $Z_L$ is the load impedance representing a known reference load or an unknown load such as a simple cavity, an ear simulator, or a human ear canal. The transmitter microphone generates a sound pressure by applying a voltage $E_{in} = \alpha_{in} P_{in}$, where $P_{in}$ is the generated sound pressure and $\alpha_{in}$ is the sensitivity. Similarly, the receiver microphone is picking up the sound pressure by measuring a voltage $E_{out} = \alpha_{out} P_{out}$. In the Thévenin equivalent acoustic circuit in Figure 5(b) $P_S$ and $Z_S$ are the Thévenin source pressure and source impedance, respectively, and $P_L$ is the load pressure.

The relationships of the Thévenin source pressure, the source impedance, and the load pressure in the Thévenin equivalent acoustic circuit, to the impedances and measurable pressures in the full acoustic circuit are given by:

$$P_S = P_{in} \frac{Z_{M2} + Z_{T2}}{Z_{M2} + Z_{T1} + Z_{T2}}$$

$$Z_S = \frac{Z_{T1} Z_{M2} + Z_{T1} Z_{T2}}{Z_{M2} + Z_{T1} + Z_{T2}} \qquad (2)$$

$$P_L = P_{out} \frac{Z_{M2} + Z_{T2}}{Z_{M2}}$$

## (a) Full acoustic circuit

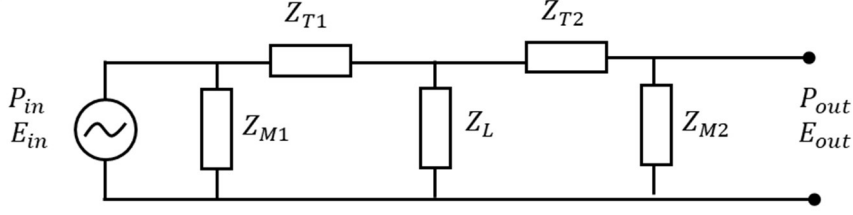

## (b) Thévenin equivalent acoustic circuit

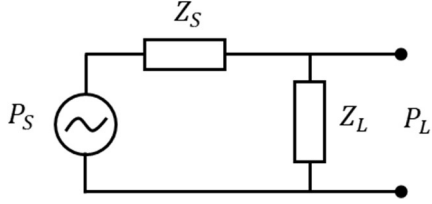

Figure 5: (a) The full acoustic circuit of the measurement setup. $Z_{T1}$, and $Z_{T2}$ are the impedances of the two waveguide tubes connecting the probe microphones and the load. $Z_{M1}$, and $Z_{M2}$ are the impedances of the impedance matching tubes inside the two probe microphones. $P_{in}$ and $E_{in}$ are the applied pressure and voltage of the transmitter probe microphone. $P_{out}$ and $E_{out}$ are the measured pressure and voltage of the receiver probe microphone. (b) The Thévenin equivalent acoustic circuit with source pressure $P_S$, source impedance $Z_S$, load pressure $P_L$, and a load impedance $Z_L$.

### A. Two-load calibration procedure with analytically calculated reference loads

From the Thévenin equivalent acoustic circuit the load pressure is given by:

$$P_L = P_S \frac{Z_L}{Z_L + Z_S} \quad (3)$$

The transfer function between the measured and the applied voltage $H_L = E_{out}/E_{in} = (\alpha_{out} P_{out})/(\alpha_{in} P_{in})$ can be found using equation ( 2 ) and equation ( 3 ):

$$H_L = K \frac{Z_L}{Z_L + Z_S} \quad (4)$$

where $K = \alpha_{out} Z_{M2}/\alpha_{in}(Z_{M2} + Z_{T1} + Z_{T2})$ is a constant determined by the sensitivities of the two probe microphones and the impedances of the full acoustic circuit. When measuring on a reference load with a known impedance an equation with two unknowns $K$ and $Z_S$ is obtained. Hence, by measuring on two known reference loads the unknowns can be estimated as:

$$K = \frac{Z_{RL2} - Z_{RL1}}{\frac{Z_{RL2}}{H_{RL2}} - \frac{Z_{RL1}}{H_{RL1}}} \qquad Z_S = \frac{H_{RL2} - H_{RL1}}{\frac{H_{RL1}}{Z_{RL1}} - \frac{H_{RL2}}{Z_{RL2}}} \qquad (5)$$

where $H_{RL1}$ and $H_{RL2}$ are the two measured transfer functions, and $Z_{RL1}$ and $Z_{RL2}$ are the impedances of the two known reference loads.

When measuring a transfer function $H_{2L}$ on an unknown load, the impedance of the unknown load $Z_{2L}$ can be found by inserting equation ( 5 ) in equation ( 4 ):

$$Z_{2L} = \frac{H_{2L}(H_{RL2} - H_{RL1})}{\frac{H_{RL1}}{Z_{RL1}}(H_{RL2} - H_{2L}) - \frac{H_{RL2}}{Z_{RL2}}(H_{RL1} - H_{2L})} \qquad (6)$$

Equation ( 6 ) is the two-load calibration procedure in its general form. If the impedance of the two known reference loads is estimated analytically by equation ( 1 ) it is referred to hereinafter as the two-load calibration procedure with analytic impedance of the reference loads.

**B.   Two-load calibration procedure with simulated reference loads**

Establishing a measurement set-up where the assumptions of equation ( 1 ) are valid is difficult to achieve in a practice at higher frequencies. As illustrated in Figure 1(b) the coupling interface of the impedance probe is intended to fit the opening of ear canals having different cross-sectional areas (typically ranging from about 25 to 75 mm$^2$). At the measurement plane excitation of the sound pressure is not done by a piston over the entire surface, and measurement is not an average over this surface, but through the two small waveguide tubes as shown in Figure 1(b). Hence, this discontinuity in cross-sectional area will excite higher order modes and introduce errors when calculating the impedance by the analytical expression given by equation ( 1 ), especially around antiresonances and at higher frequencies [28–30], see also Figure 3(a). It is therefore desirable to use a more exact estimate of the reference load impedance from finite element simulation taking into account the exact geometry of the reference load cavity and waveguide tubes, as shown in Figure 4. Simulated impedances of all reference loads are therefore calculated and used to substitute $Z_{RL1}$ and $Z_{RL2}$ in the two-load calibration procedure of equation ( 6 ), which in the following will be referred to as the two-load calibration procedure with simulated impedances of the reference loads. Note that $Z_{2L}$ as calculated by equation ( 6 ) then also expresses the input impedance at the orifice of the measuring waveguide tube rather than the average impedance over the full surface.

### C.    Multi-load calibration procedure with simulated reference loads

The two-load calibration procedure can be improved by measuring on more than two known reference loads, yielding an overdetermined set of equations for finding the unknowns in equation ( 5 ) which may be solved by the least square method as developed by Nørgaard et.al. [29]. However, this requires transfer function measurements on several reference loads to be carried out at any measurement session. It may be quite cumbersome to perform in an in-vivo study where several human factors contribute to the difficulties in obtaining reliable measurements and where measurements often have to be repeated. A larger study may take place over several days or weeks. A simpler calibration procedure is therefore developed here, where an average two-load impedance and an uncertainty is estimated. This procedure makes repetitive use of Eq. (6) by considering two reference loads at a time and is as such a simple extension of the traditional two-load method.

By writing the two-load calibration procedure as:

$$Z_{2L_{i,j}} = \frac{H_{2L}(H_{RL_j} - H_{RL_i})}{\frac{H_{RL_i}}{Z_{RL_i}}(H_{RL_j} - H_{2L}) - \frac{H_{RL_j}}{Z_{RL_j}}(H_{RL_i} - H_{2L})} \qquad (7)$$

where $Z_{RLi}$ and $Z_{RLj}$ are the impedances of the known reference loads number $i$ and $j$ respectively, and $H_{RLi}$ and $H_{RLj}$ are the measured transfer functions on the known reference loads number $i$ and $j$ respectively. Here $i = 1 \ldots n-1$ and $j = i+1 \ldots n$, and $n$ is the number of known reference loads. This yields $\frac{1}{2}n(n-1)$ estimates of the unknown load $Z_{2L}$ from which an average impedance $Z_{ML}$ and standard deviation $\sigma_{Z_{ML}}$ can be estimated:

$$Z_{ML} = \frac{2}{n(n-1)} \sum_{i=1}^{n-1} \sum_{j=i+1}^{n} Z_{2L_{i,j}} \tag{8}$$

$$\sigma_{Z_{ML}} = \sqrt{\frac{2}{n(n-1)-2} \sum_{i=1}^{n-1} \sum_{j=i+1}^{n} \left(Z_{ML} - Z_{2L_{i,j}}\right)^2} \tag{9}$$

In the following equation ( 8 ) will be referred to as the multi-load calibration procedure, and equation ( 9 ) the estimated standard deviation. Since the $\frac{1}{2}n(n-1)$ estimates of $Z_{2L_{i,j}}$ are not fully uncorrelated, equation ( 9 ) is strictly speaking not the standard deviation, but a good estimate of the standard deviation.

### D.   Multi-load with simulated reference loads post processed, semi multi-load

As shown in the following the multi-load calibration procedure can be applied as post processing to correct any two-load measurement already obtained using only two loads. This method will be referred to as the semi multi-load calibration procedure.

A one-time multi-load calibration procedure performed on the two selected reference loads yields $\frac{1}{2}(n-1)(n-2)$ new estimates of each of their impedances, from which averages are estimated using the multi-load calibration procedure in equation ( 8 ):

$$Z'_{RL1} = \frac{2}{(n-1)(n-2)} \sum_{i=2}^{n-1} \sum_{j=i+1}^{n} Z_{2L_{i,j}} \quad (10)$$

and similarly, for $Z'_{RL2}$.

The new impedance estimates of the two known reference loads $Z'_{RL1}$ and $Z'_{RL2}$ are based on both simulation and measurements, and they can be used to improve the two-load calibration procedure by replacing $Z_{RL1}$ and $Z_{RL2}$ in equation (6).

Another way to use the new estimates ($Z'_{RL1}$ and $Z'_{RL2}$) to improve the two-load calibration procedure is the following: If the change in the known reference load impedances $\Delta Z_{RL1} = Z'_{RL1} - Z_{RL1}$ and $\Delta Z_{RL2} = Z'_{RL2} - Z_{RL2}$ are relatively small $\left|\frac{\Delta Z_{RL1}}{Z_{RL1}}\right|, \left|\frac{\Delta Z_{RL2}}{Z_{RL2}}\right| \ll 1$, the multi-load calibration procedure can be approximated by:

$$Z_{SML} = \frac{\partial Z_{2L}}{\partial Z_{RL1}} \Delta Z_{RL1} + \frac{\partial Z_{2L}}{\partial Z_{RL2}} \Delta Z_{RL2} + Z_{2L} \quad (11)$$

called the semi multi-load calibration procedure. Here the two differential coefficients are assumed linearly independent. By approximating:

$$\frac{\Delta Z_{RL1}}{Z_{RL1}} \cong \frac{\Delta Z_{RL2}}{Z_{RL2}} \cong \frac{1}{2}\left(\frac{\Delta Z_{RL1}}{Z_{RL1}} + \frac{\Delta Z_{RL2}}{Z_{RL2}}\right) \quad (12)$$

it can be shown from equations (6) and (12) that the impedance $Z_{SML}$ can be written as:

$$Z_{SML} \cong \frac{1}{2}\left(\frac{Z'_{RL1}}{Z_{RL1}} + \frac{Z'_{RL2}}{Z_{RL2}}\right) Z_{2L} \quad (13)$$

where $Z_{2L}$ is the unknown load calibrated by the two-load calibration procedure in equation (6). $Z'_{RL1}$ and $Z'_{RL2}$ are the known reference loads calibrated by the multi-load calibration procedure in equation (10), and $Z_{RL1}$ and $Z_{RL}$ are the known reference loads used in the two-load calibration procedure, estimated either by the analytic expression in equation (1) or by simulation.

The relation between semi multi-load impedance and two-load calibrated impedance as expressed in equation ( 13 ) is derived in Appendix. It increases the accuracy of measurements already calibrated by the two-load calibration procedure of equation ( 6 ). Therefore, the advantage of the semi multi-load calibration procedure, is that it is as simple to use in practice as the two-load calibration procedure, but it still has high accuracy close to the full multi-load calibration procedure, see Figure 9.

## V. COMPARISON OF CALIBRATION PROCEDURES

In the following, the two-load and the multi-load calibration procedures are compared based on impedance measurements on the cavity V1006.

### A. Two-load calibration procedures

Figure 6 shows the measured impedance of cavity V1006 calibrated by the two-load calibration procedure using analytic and simulated impedances of the reference loads in the orange and green curve, respectively. The two reference loads used for calibration were V1285 and V200. A finite element simulation of the impedance of V1006 is shown as reference in blue. Below 4 kHz a good agreement is observed between the two measured and the simulated response (the orange and green versus the blue curve). As the frequency approaches the first antiresonance at 6 kHz (on the blue curve), corresponding to the frequency where a quarter wavelength becomes equal to the length of cavity V1006 (15.81 mm), the measured responses start to deviate. This is most easily observed in the magnitude, Figure 6(a). The measured response calibrated using analytic expression for the impedance of the reference loads deviates more and estimates a lower and less well-defined quarter wavelength resonance than when using the simulated expression, (the orange curve versus the green curve). Above the half wavelength resonance at 10.9 kHz large deviations are observed. By studying the simulated impedance of the large reference load cavity V1285 as shown in Figure 4, it is clear that the antiresonances of V1285 at 6 kHz and 18.5 kHz creates large errors around the same frequencies in the measurement of V1006 calibrated using analytic expression for the impedance of the reference loads, (the orange curve in Figure 6(a)). It is also noted that both two-load calibration procedures generate a negative resistance around the half wavelength resonance, see Figure6(b).

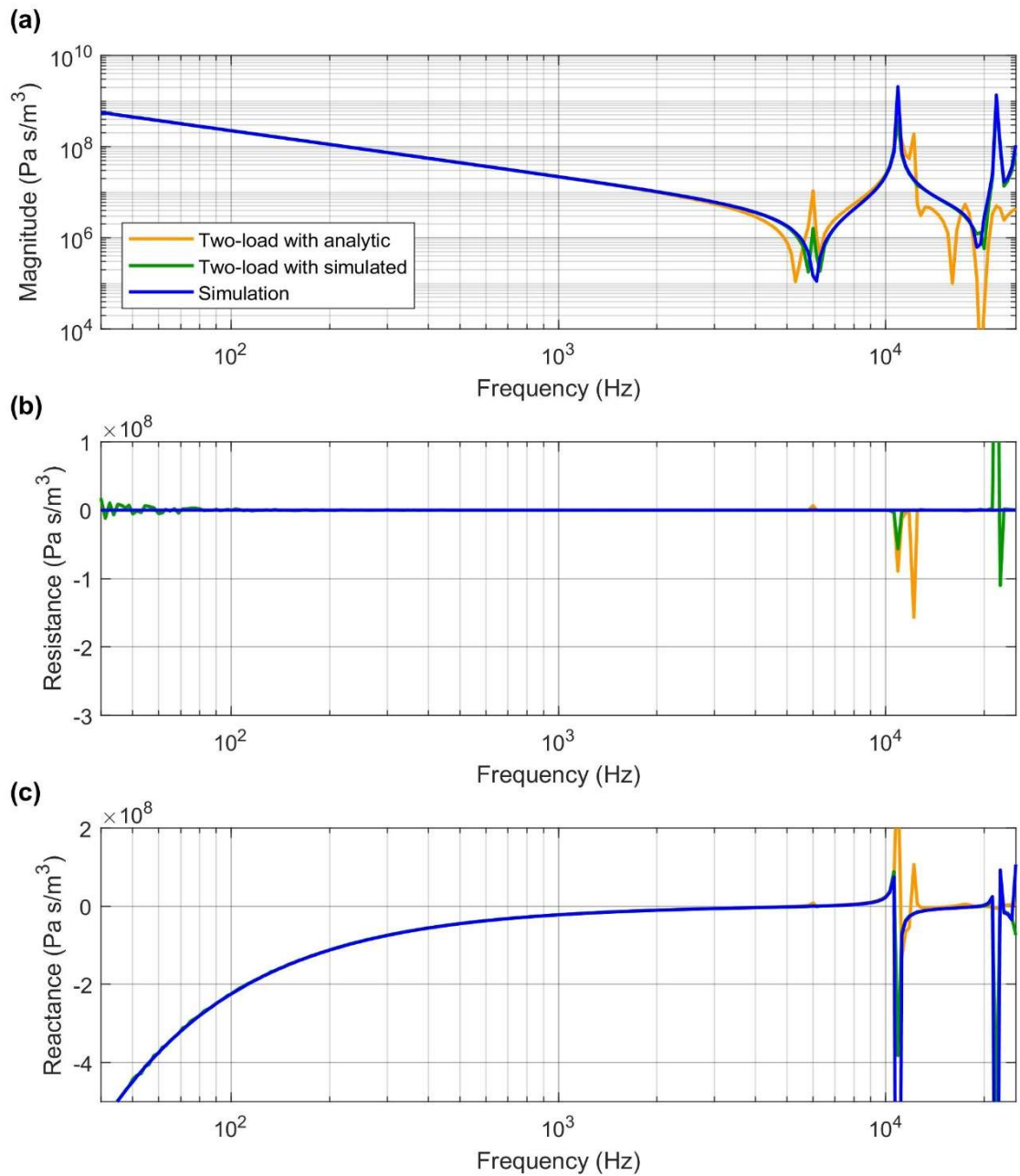

Figure 6: Input impedance of the cavity V1006. Measurements calibrated by the two-load calibration procedure using analytic and simulated impedances of the reference loads are shown in orange and green, respectively. A finite element simulation of the impedance is shown in blue.

### B.  Multi-load calibration procedures

Figure 7 shows the impedance of reference load V1006 (see Table I) calibrated by the multi-load and the semi multi-load calibration procedure. The multi-load curve has error bars with a size of three times the standard deviation (equation ( 9 )). The standard deviation, or uncertainty, can be minimized by selecting the proper reference loads. Large uncertainties may be introduced if the impedances of the reference loads are approximately the same, or if both are either much smaller or much larger than the unknown load. Since the impedance of the unknown load may change considerably over the frequency range, dividing the frequency range into two using different sets of reference loads is often a good choice. Large uncertainties may also be introduced if the reference loads have resonances within the frequency range of interest, because at resonances the measured and the simulated impedances are difficult to determine with low uncertainty. The multi-load curve in Figure 7 is therefore calibrated using load V100, V200, V759, and V1285 in the frequency range from 35 Hz to 3 kHz, and using V40, V65, V100 and V200 in the frequency range from 3 kHz to 25 kHz. The semi multi-load curve in Figure 7 is calibrated using V759 and V1285 in the frequency range from 35 Hz to 3 kHz, and V65 and V200 in the frequency range from 3 kHz to 25 kHz. The smaller loads are best in the high frequency range since they do not have resonances in this range as the larger loads have. The larger loads are best in the low frequency range since they have lower impedances that are typically more comparable to the unknown load.

Since the multi-load calibration procedure is the average of all the possible two-load combinations, the error bars of the multi-load curve in Figure 7 correspond to the range of impedance estimates when using the two-load calibration procedure with simulated impedance of the reference loads.

In Figure 7(a) the magnitude of the impedance is plotted, and it shows that the multi-load and the semi multi-load curve are almost identical. The error bars are also very narrow, meaning that magnitude estimates using the two-load calibration procedure with simulated impedance of the reference loads are very close to the magnitude of the multi-load calibration procedure.

Figure 7(b) shows that the resistance of the multi-load and the semi multi-load curve are very similar except at the lowest frequencies, and that they are both positive up to at least 20 kHz. However, the error bars also show that impedance estimates using the two-load calibration procedure can easily yield a negative resistance in some of the frequency range, which is not physically possible in a passive simple cylinder such as V1006.

The problem with negative resistance is also seen when applying the two-load calibration procedure to more complex cavities such as the human ear canal. When using the multi-load or the semi multi-load calibration procedure this problem diminishes, and a positive resistance is usually obtained.

Figure 7(c) shows that the reactance of the multi-load and semi multi-load curves are very similar. The error bars are also narrow up to at least 20 kHz (except close to the half wavelength resonance), meaning that reactance estimates using the two-load calibration procedure with simulated impedance of the reference loads are very close to the reactance of the multi-load calibration procedure.

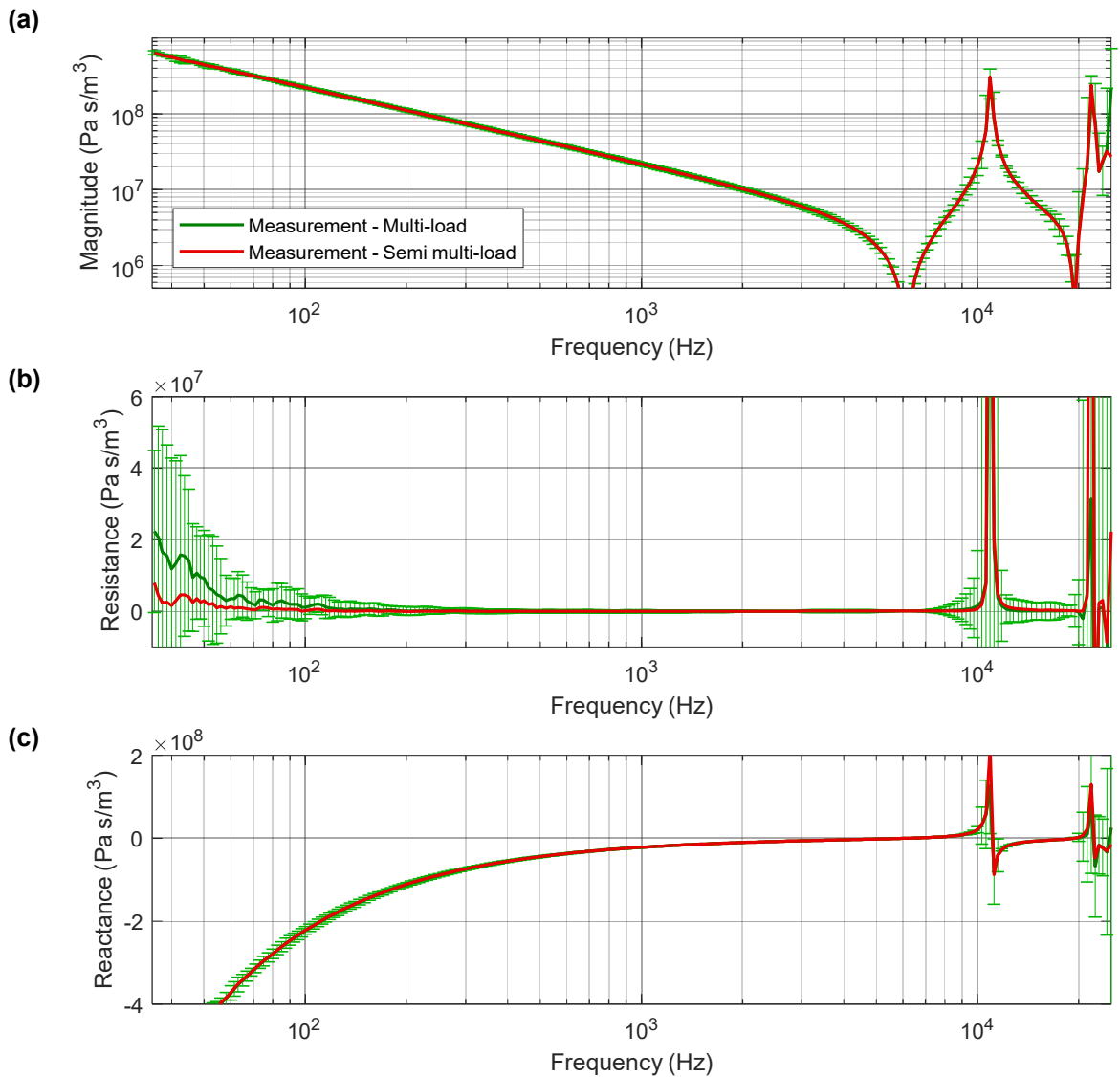

Figure 7: Input impedance of the cavity V1006. Measurements calibrated by the multi-load and the semi multi-load calibration procedure are shown in green and red, respectively. The error bars in green show three times the standard deviation of the multi-load calibration procedure and correspond to the range of impedance estimates when using the two-load calibration procedure with simulated impedance of the reference loads.

# VI. VERIFICATION OF MULTI-LOAD CALIBRATION PROCEDURES

In the following section, the accuracy of the multi-load and the semi multi-load calibration procedures are verified, partly by comparing a measurement of the input impedance of the IEC711 coupler calibrated by these procedures to a detailed simulation, and partly by examining the standard deviation of the multi-load calibration procedure.

Figure 8 shows measured input impedance of the IEC711 coupler calibrated by the multi-load calibration procedure and the semi multi-load calibration procedure in the green and red curves, respectively. Finite element simulation including viscothermal loss of the input impedance of the IEC711 coupler is shown by the blue curves. For the multi-load calibration procedure, reference loads V200, V759, V1006, and V1285 are used in the frequency range from 35 Hz to 3 kHz, and V40, V65, V100, and V200 are used from 3 kHz to 25 kHz. The crossover point at 3 kHz between the two frequency ranges is chosen so that the standard deviation is minimized. For the semi multi-load calibration procedure reference load V759 and V1285 is used in the frequency range from 35 Hz to 3 kHz, and V65 and V200 is used from 3 kHz to 25 kHz. As the semi multi-load is applied as post-processing, it is obvious to use as many reference loads s as possible. However, in order to determine if a minimum number of reference loads still provides satisfactory results, two sets of only 2 reference loads of appropriate sizes have been selected here.

The multi-load and the semi multi-load calibration procedures yield very similar estimates of the impedance of the IEC711 coupler. The problem with negative resistance when using the two-load calibration procedure can also be observed in Figure 8(b). The error bars show that statistically some two-load combinations will yield a negative resistance below app. 150 Hz and beyond 20 kHz.

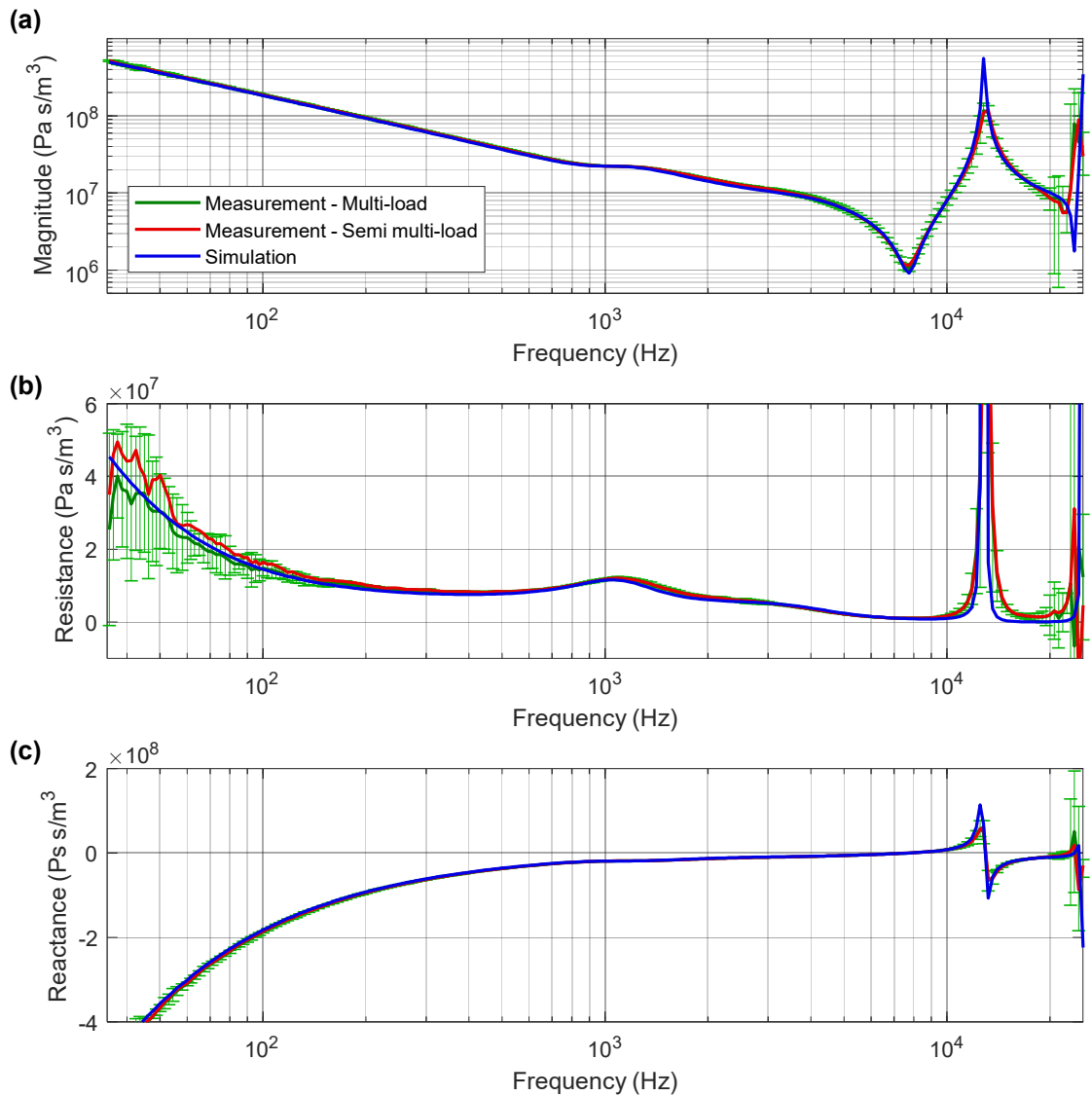

Figure 8: Input impedance of the ear simulator (IEC711 coupler). Measurements calibrated by the multi-load and the semi multi-load calibration procedure are shown in green and red, respectively. Finite element simulation including viscothermal loss is shown in blue. The error bars in green show three times the standard deviation of the multi-load calibration procedure and correspond to the range of impedance estimates when using the two-load calibration procedure with simulated impedance of the reference loads.

Figure 9 shows the relative difference between the simulated and the measured input impedance of the IEC711 coupler using the multi-load calibration procedure (green curves) and the semi-load calibration procedure (red curves):

$$\Delta_{Rel} = \left| \frac{Z_{sim} - Z_{meas}}{Z_{meas}} \right| \quad (14)$$

where $Z_{sim}$ is the simulated impedance, and $Z_{meas}$ is the measured impedance calibrated by the multi-load or the semi multi-load calibration procedure. Three times the standard deviation of the measured impedance calibrated by the multi-load calibration procedure is also shown in black. The frequency axis in Figure 9 ends at 20 kHz and not at 25 kHz (as in Figure 8) since the calibration procedures are only validated up to 20 kHz. In Figure 9(a) the magnitude is plotted, and it shows that $\Delta_{Rel}$ for the multi-load and the semi multi-load calibration procedures and for the standard deviation are below ~10% in most of the frequency range up to 20 kHz, except close to resonances. This is also the case for the reactance as shown in Figure 9(c). However, for the resistance as shown in Figure 9(b), $\Delta_{Rel}$ and the standard deviation especially close to and beyond the half wavelength resonance are much higher. This is probably because the resistance is much smaller than the reactance and therefore difficult to measure, with low uncertainty, i.e., the reactance in Figure 8(c) is about ten times higher than the resistance in Figure 8(b).

Since the measured input impedance magnitude calibrated by the multi-load and semi multi-load calibration procedure (using two sets of four and two sets of two reference loads, respectively), has a relatively small discrepancy compared to a detailed simulation, and since the standard deviation of the multi-load calibration procedure is small, these procedures seem to give a reliable estimate of the input impedance magnitude of a complex geometrical structure as the IEC711 coupler, in the full audio bandwidth up to 20 kHz.

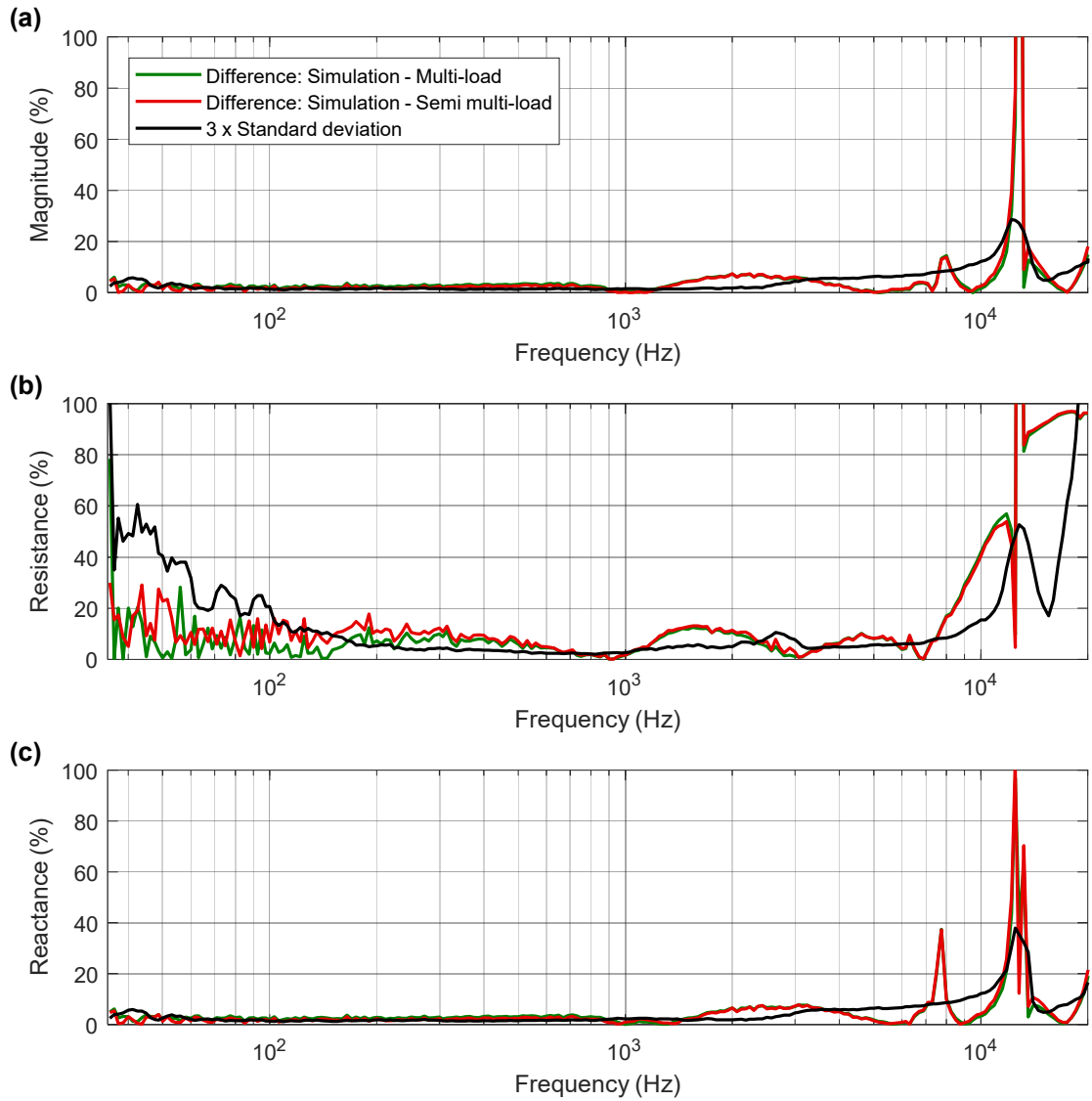

Figure 9: Relative difference $\Delta_{Rel}$ between the simulated and the measured input impedance (equation ( 14 )) of the IEC711 coupler in two cases. The measurements calibrated by the multi-load and the semi multi-load calibration procedures are shown in green and red, respectively. The black curve shows three times the standard deviation of the measured impedance calibrated by the multi-load calibration procedure.

## VII. HUMAN EAR CANAL MEASUREMENTS

To further validate the multi-load calibration procedure a measurement on a male adult subject with normal hearing, aged 41, was prepared. In order to obtain an optimal seal and fit, when applying the impedance probe assembly of Figure 1 to the ear canal, an individual ear mould was made as shown in Figure 10. The ear mould has two holes for connecting the acoustic waveguide tubes of the impedance probe assembly. The holes are placed with the same spacing as was used when connecting the impedance probe to the reference load cavities. The measuring plane of the ear mould is positioned between the first and second bend of the ear canal.

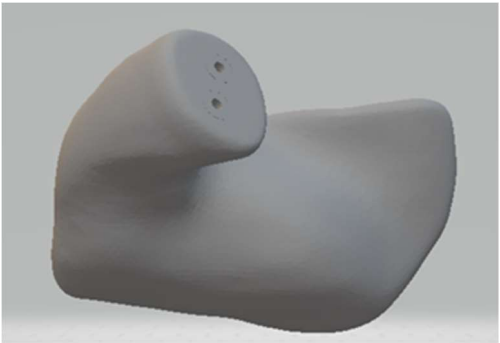

Figure 10: Geometry of the individual ear mould used when measuring the impedance of one human subject. The ear mould has two holes for connecting the acoustic waveguide tubes of the impedance probe assembly, see Figure 1. The measuring plane of the ear mould was positioned between the first and second bend of the ear canal.

The measured input impedance of the adult subject is shown in Figure 11. The green curve was calibrated by the multi-load calibration procedure using reference loads V200, V759, V1006, and V1285 from 35 Hz to 3 kHz, and V40, V65, V100, and V200 from 3 kHz to 25 kHz, and the error bars in green show three times the standard deviation. The red curve was calibrated by the semi multi-load calibration procedure using V759 and V1285 from 35 Hz to 3 kHz, and V65 and V200 from 3 kHz and 25 kHz.

Figure 11 shows that the multi-load and the semi multi-load curves yield very similar estimates of the impedance in the full frequency range, and that the standard deviation is very low up to 20 kHz, quite similar to the impedance measured on the IEC711 coupler in Figure 8. Both the multi-load and the semi multi-load calibration procedure is therefore expected to provide accurate and reliable results for measurements of the input impedance magnitude of the human ear in the full audio bandwidth up to 20 kHz. The semi multi-load calibration procedure is subsequently applied in a larger in-vivo impedance study on human ears[20].

By studying the impedance in Figure 11 several characteristic features of the human ear may be observed. In the frequency range below 800 Hz where the eardrum is primarily stiffness controlled the magnitude of the subject, as well of the IEC711 coupler of Figure 8, follows a straight line with no further roll off towards 35 Hz, indicating that no leakage is introduced. Towards the mid-frequencies around 1300 Hz the response starts to rise. This may identify the first resonance of the middle-ear being observed through the eardrum, where the response changes from being stiffness controlled to mass controlled. The slope of the line gives an indication of the damping in the middle ear and from the flat part of the raised level above 1300 Hz the ear canal volume between the measuring plane and the eardrum may be approximated. At approx. 5.5 kHz the antiresonance corresponds to a quarter wavelength resonance in the ear canal between the measuring plane and the eardrum. At approx. 9.1 kHz half a wavelength resonance occurs and so forth. The standing wave pattern in the ear canal can be observed to above 20 kHz.

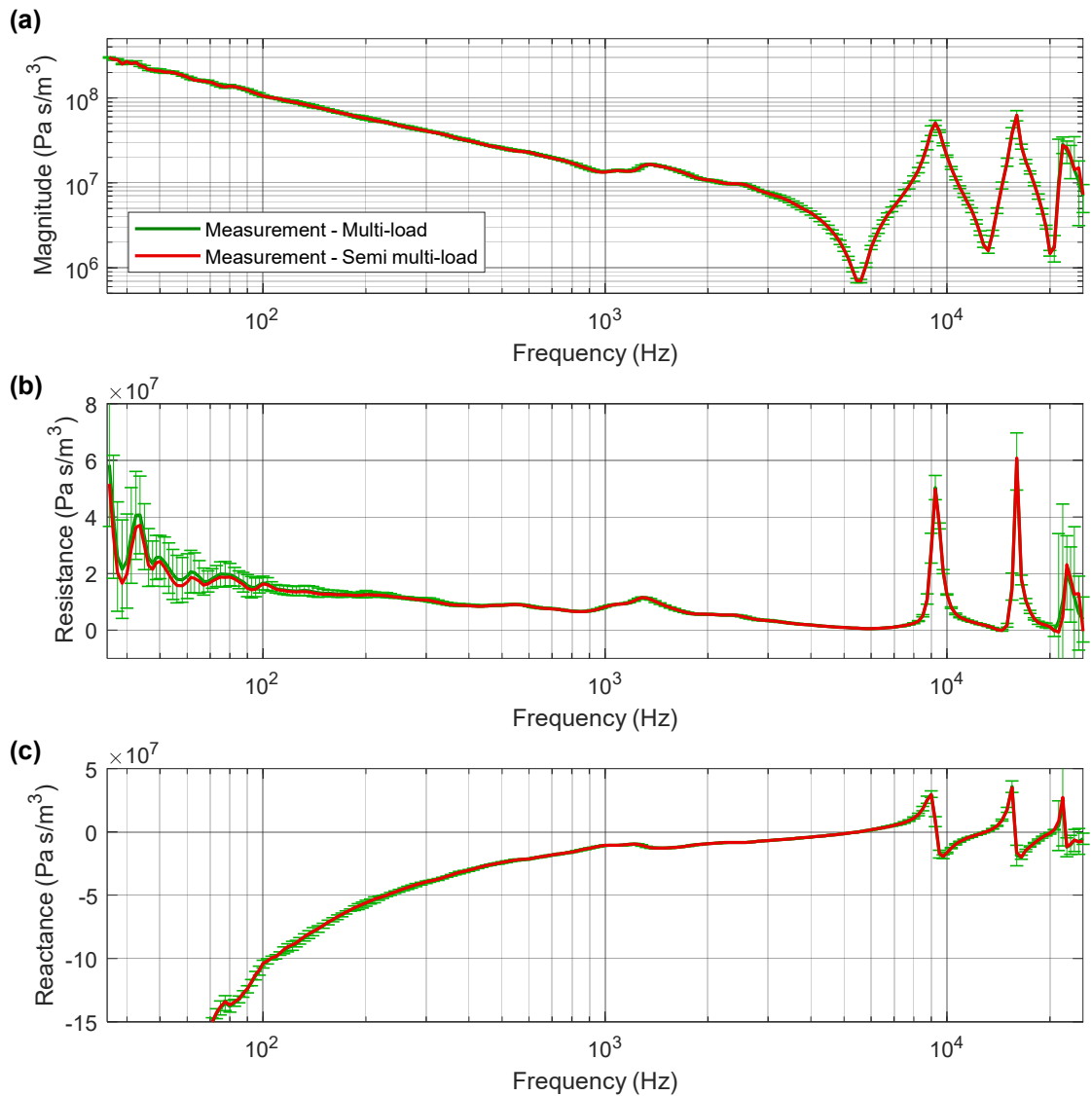

Figure 11: Measured input impedance of a human ear canal calibrated by the multi-load and the semi multi-load calibration procedure in green and red, respectively. The errorbars in green shows three times the standard deviation of the multi-load calibration procedure and correspond to the range of impedance estimates when using the two-load calibration procedure with simulated impedance of the reference loads.

# VIII. CONCLUSION

Today impedance measurements of the human ear are only well covered below 8-10 kHz. An important factor contributing to the difficulties in obtaining accurate measurements at higher frequencies is that the wavelength of sound becomes close to the dimensions within the ear canal and concha, which causes resonances and cross modes to be excited. Traditional impedance measurement methods, as the two-load calibration procedure investigated here, often become inaccurate when the frequency approaches antiresonances, which may occur around 5-6 kHz in a typical human ear canal. Therefore, the purpose of this study has been to develop a reliable wideband impedance measurement method suitable for a larger scale in-vivo study of the human ear.

First, the two-load method was refined by replacing the analytic expressions for the known impedance of the reference loads by detailed finite element simulations. This improved accuracy and widened the frequency range, since the simulated impedance includes all details of the geometry of the reference loads in contrast to the analytic expression. However, the two-load calibration procedures still tend to generate a non-physical negative resistance, which in particular made it difficult to estimate the correct impedance level at resonances.

The refined two-load calibration method was therefore extended to a multi-load method using several reference loads from which an impedance average of all possible two-load combinations was found. Another improvement consisted of dividing the frequency range into two ranges, each using a different set of reference loads. By using smaller reference loads at higher frequencies, it was possible to move undesirable effects of resonances and possible cross modes beyond the frequency range of interests and thereby extend the frequency range of the calibration method.

The multi-load calibration procedure generally solved the problem with negative resistance and when applied to an ear simulator, the IEC711 coupler, it showed low standard deviation on the impedance magnitude and good agreement compared to a detailed simulation in the full audio bandwidth up to 20 kHz.

The multi-load calibration procedure was further processed, to create the semi multi-load calibration procedure, which allows to apply the method as postprocessing. It was done to avoid calibration of all reference loads during each measurement session, thus making the method more efficient to use in practice. Results obtained on a simple cavity of 1006 mm$^3$ and the IEC711 coupler provided very similar estimates of the impedance with both calibration procedures, even beyond 20 kHz.

Finally, the multi-load and the semi multi-load calibration procedure was applied to an impedance measurement on a human ear canal. Again, very similar impedance results and a low standard deviation was obtained in the full audio bandwidth and some characteristic features of the human ear that were observed from the human impedance were discussed. Both calibration procedures were found suitable for measuring wideband impedance of the human ear. The semi multi-load calibration procedure is more efficient to use in practise and has subsequently been applied in a larger in-vivo study.

## ACKNOWEDGEMENT

Thanks to the transducer team at Brüel & Kjær for suggestions and comments to the manuscript.

## APPENDIX

This appendix contains derivation of the semi multi-load calibration procedure of equation ( 13 ).

The semi multi-load calibration procedure is defined by:

$$Z_{SML} = \frac{\partial Z_{2L}}{\partial Z_{RL1}}\Delta Z_{RL1} + \frac{\partial Z_{2L}}{\partial Z_{RL2}}\Delta Z_{RL2} + Z_{2L} \qquad (a15)$$

The two differential coefficients in equation ( a15 ) are assumed linearly independent and given by:

$$\frac{\partial Z_{2L}}{\partial Z_{RL1}} = Z_{2L}\frac{1}{\frac{H_{RL1}}{Z_{RL1}}(H_{RL2} - H_{2L}) - \frac{H_{RL2}}{Z_{RL2}}(H_{RL1} - H_{2L})}\frac{H_{RL1}}{Z_{RL1}^2}(H_{RL2} - H_{2L}) \qquad (a16)$$

$$\frac{\partial Z_{2L}}{\partial Z_{RL2}} = -Z_{2L}\frac{1}{\frac{H_{RL1}}{Z_{RL1}}(H_{RL2} - H_{2L}) - \frac{H_{RL2}}{Z_{RL2}}(H_{RL1} - H_{2L})}\frac{H_{RL2}}{Z_{RL2}^2}(H_{RL1} - H_{2L}) \qquad (a17)$$

Inserting equation ( a16 ) and ( a17 ) in equation ( a15 ) yields:

$$Z_{SML} = Z_{2L}\left(\frac{\Delta Z_{RL1}}{Z_{RL1}}\frac{1}{\frac{H_{RL1}}{Z_{RL1}}(H_{RL2} - H_{2L}) - \frac{H_{RL2}}{Z_{RL2}}(H_{RL1} - H_{2L})}\frac{H_{RL1}}{Z_{RL1}}(H_{RL2} - H_{2L})\right.$$

$$\left. -\frac{\Delta Z_{RL2}}{Z_{RL2}}\frac{1}{\frac{H_{RL1}}{Z_{RL1}}(H_{RL2} - H_{2L}) - \frac{H_{RL2}}{Z_{RL2}}(H_{RL1} - H_{2L})}\frac{H_{RL2}}{Z_{RL2}}(H_{RL1} - H_{2L}) \right. \qquad (a18)$$

$$\left. + 1 \right)$$

By using the approximation $\frac{\Delta Z_{RL1}}{Z_{RL1}} \cong \frac{\Delta Z_{RL2}}{Z_{RL2}} \cong \frac{1}{2}\left(\frac{\Delta Z_{RL1}}{Z_{RL1}} + \frac{\Delta Z_{RL}}{Z_{RL}}\right)$, equation( a18 ) can be written as:

$$Z_{SML} \cong Z_L\left(\frac{1}{2}\left(\frac{\Delta Z_{RL1}}{Z_{RL1}} + \frac{\Delta Z_{RL2}}{Z_{RL2}}\right) + 1\right) \qquad (a19)$$

By using the definitions $\Delta Z_{RL1} = Z'_{RL} - Z_{RL1}$ and $\Delta Z_{RL} = Z'_{RL} - Z_{RL}$ equation ( a19 ) can be written as:

$$Z_{SML} \cong \frac{1}{2}\left(\frac{Z'_{RL1}}{Z_{RL}} + \frac{Z'_{RL2}}{Z_{RL}}\right)Z_{2L} \qquad (a20)$$